# Imaging nuclear spins weakly coupled to a probe paramagnetic center

*Abdelghani Laraoui[1], Daniela Pagliero[1], Carlos A. Meriles[1]*

[1]Dept. of Physics, CUNY-City College of New York, New York, NY 10031, USA.

**Abstract**

Optically-detected paramagnetic centers in wide-bandgap semiconductors are emerging as a promising platform for nanoscale metrology at room temperature. Of particular interest are applications where the center is used as a probe to interrogate other spins that cannot be observed directly. Using the nitrogen-vacancy center in diamond as a model system, we propose a new strategy to determining the spatial coordinates of weakly coupled nuclear spins. The central idea is to label the target nucleus with a spin polarization that depends on its spatial location, which is subsequently revealed by making this polarization flow back to the NV for readout. Using extensive analytical and numerical modeling, we show that the technique can attain high spatial resolution depending on the NV lifetime and target spin location. No external magnetic field gradient is required, which circumvents complications resulting from changes in the direction of the applied magnetic field, and considerably simplifies the required instrumentation. Extensions of the present technique may be adapted to pinpoint the locations of other paramagnetic centers in the NV vicinity or to yield information on dynamical processes in molecules on the diamond surface.

Nuclear magnetic resonance (NMR) excels in its ability to probe non-destructively the structure and dynamics of molecular moieties without the need for long-range order, but the low detection sensitivity inherent to inductive detection limits this technique to ensemble measurements. Magnetic resonance force microscopy (MRFM) was introduced two decades ago precisely to circumvent this problem[1]. The idea was to leverage on the extreme sensitivity and magnetic field gradients of MRFM probes to discriminate between signals from individual nuclear sites in a molecule so as to image its structure with atomic resolution. Despite enormous progress[2], however, this latter goal has proven exceedingly difficult, mainly due to the minute ratio between the spin and thermal energies, even at the lowest temperatures possible today.

A more recent route to nanoscale magnetic resonance builds on the properties of select paramagnetic centers in solid-state matrices. At the core of this form of sensing is a singular combination of transition rules and decay rates between electronic states, responsible for the up-conversion of spin-flip-induced energy differences into, e.g., changes in the center's fluorescence. Case systems presently under intense study are the di-vacancy and silicon-vacancy centers in SiC[3-5], rare-earth ions in garnets[6,7], substitutional phosphorous[8] and bismuth[9] in silicon, and the nitrogen-vacancy (NV) center in diamond[10]. Common to these systems is the ability to individually initialize, manipulate, and readout the paramagnetic center spin via a combination of microwave (mw) and optical (or electrical) pulses, often under ambient conditions. These unique features are being presently exploited to probe other spins in the paramagnetic center vicinity that cannot be directly initialized or interrogated. For example, nitrogen-vacancy (NV) centers engineered near the diamond surface have been used to detect small ensembles of protons from organic films or fluids in contact with the crystal surface[11,12]. Subsequent experiments[13] have demonstrated the detection of other spin species including $^{19}$F, and $^{31}$P, with more recent studies reporting the observation of individual $^1$H and $^{29}$Si spins[14,15]. In this light, schemes designed to determine the spatial location of individual nuclear (or electronic) spins with atomic resolution gain particular interest as they are likely to become a key ingredient in, e.g., the characterization of the structure and dynamics of individual molecules adsorbed onto the center's solid-state host.

Here we introduce an imaging protocol that builds on the hyperfine-induced gradient to selectively mediate the transfer and retrieval of spin polarization between the paramagnetic center and neighboring nuclei. In many ways, our approach follows the principles of radar technology where a probe 'signal'—in the form of spin polarization—scouts the vicinity in search for a 'target'. In our scheme, the location of the target—i.e., a nuclear spin—is determined as the 'reflected' polarization travels back to the source. The target's spatial coordinates are reconstructed from the angle where the reflection took place and the round trip time, here encoded in the pulse phases and duration of our magnetic resonance sequence. Both the range and spatial resolution depend on the coherence lifetime of the paramagnetic center, which varies from one spin system to another. For concreteness, we consider the particular case of an NV center interacting with distant (i.e., 0.5 nm or more) nuclear spins. We find that the spatial resolution—itself a complex function of the nuclear spin location and NV lifetimes—is sufficient to image small NV/$^{13}$C clusters without the need for external magnetic field gradients. Since the applied magnetic field remains unchanged, our approach circumvents complications arising from NV level mixing and can be extended to high field (where changing the relative sample orientation is often impractical). Therefore, these ideas complement recent experimental work where the coordinates of individual proton spins from molecules on the diamond surface are determined by systematically altering the field direction in the presence of dynamical decoupling[16].

To more formally introduce our imaging scheme, we start by considering the individual NV-$^{13}$C spin pair highlighted in Fig. 1a. Here we take a reference frame centered at the NV—a system with spin number $S = 1$—and use the spherical coordinates $(r, \theta, \varphi)$ to indicate the location of the $^{13}$C nucleus—featuring a spin number $I = 1/2$. We assume that the NV and z-axis are parallel and co-linear with an externally applied magnetic field $B_0$. The system Hamiltonian is then given by

$$H = \Delta S_z^2 - \gamma_S B_0 S_z + A_{z\parallel} S_z I_z + A_{z\perp} S_z I_\varphi - \gamma_I B_0 I_z , \qquad (1)$$

where, as usual, $\Delta = 2\pi \times 2.87$ GHz denotes the NV zero-field splitting, $\gamma_S$ ($\gamma_I$) is the NV ($^{13}$C) gyromagnetic ratio, $S_z$ and $I_z$ are the z-projections of the NV and $^{13}$C spin operators, and $I_\varphi \equiv I_x \cos\varphi + I_y \sin\varphi$, with $\varphi$ denoting the nuclear spin azimuthal angle (Fig. 1a). For a distant carbon, the NV-$^{13}$C coupling is dipolar in nature and the hyperfine constants take the form $A_{z\parallel} = k\,(3\cos^2\theta - 1)/r^3$ and $A_{z\perp} = 3k\,\cos\theta\,\sin\theta/r^3$ where $k \equiv \mu_0 \gamma_{NV} \gamma_C \hbar/(4\pi)$, $\mu_0$ is the vacuum permeability, and $\hbar$ is Planck's constant divided by $2\pi$. In a typical magnetic resonance experiment, microwave (mw) pulses induce transitions between two selected states of the NV ground state triplet thus rendering the NV a two-level system with characteristic frequency $\omega_S$. The system evolution can then be described via the simplified Hamiltonian

$$H^* = A_{z\parallel} S_z I_z + A_{z\perp} S_z e^{-i\omega_I t I_z} I_\varphi e^{i\omega_I t I_z} , \qquad (2)$$

where the star denotes a transformation to the doubly rotating frame at the NV and $^{13}$C Larmor frequencies, and we have ignored rapidly fluctuating terms. Unless otherwise noted, we will assume that the hyperfine coupling is weak compared to the nuclear Zeeman field, i.e., $A_{z\parallel}, A_{z\perp} \ll \omega_I \equiv \gamma_I B_0$.

To scout for nuclear spin targets we make use of the INEPT-like[17] polarization transfer protocol in Fig. 1b comprising a train of mw pulses on the NV spin (a 'CPMG train') followed by an interval of free evolution and a radio-frequency pulse resonant with the $^{13}$C frequency (see below). For reasons that will be apparent shortly, we assume that all mw pulses induce transitions between the $m_S = \pm 1$ states, possible via the use of composite pulses[18] or multi-frequency excitation[19]. Throughout the protocol, both the duration of the train and the free evolution interval—respectively denoted as $t_1, t_2$—are systematically increased; in the case of $t_1$, the increments are in units of $\tau$—coincident with half the $^{13}$C Larmor period—and accompanied by the addition of a new $\pi$-pulse at each step. To best appreciate the combined effect of the pulse sequence we first note that over a unit $\tau/2$—$\pi$—$\tau$—$\pi$—$\tau/2$ of the CPMG train the effective system evolution can be calculated from the average Hamiltonian[20]

$$H_1^* \approx \int_0^{\tau/2} H^*(t)\, dt - \int_{\tau/2}^{3\tau/2} H^*(t)\, dt + \int_{3\tau/2}^{2\tau} H^*(t)\, dt = \frac{2}{\pi} A_{z\perp} S_z I_\varphi \ . \qquad (3)$$

independent of $A_{z\parallel}$ (see Section S1 of the Supplementary Material). The alternating sign in the sum above results from the cumulative effect of the $\pi$-pulses, here inverting the populations between the $m_S = \pm 1$ states and consequently mapping $S_z$ into $-S_z$ each time. Conversely, the effective Hamiltonian governing the $t_2$ interval is given by

$$H_2^* \approx \int_0^{t_2} H^*(t)\, dt = A_{z\parallel} S_z I_z \ , \qquad (4)$$

independent of $A_{z\perp}$.

We now calculate the system evolution, which, for simplicity, we restrict to the case where $t_1$ and $t_2$ take the (optimum) values $t_1^{opt} \equiv \pi^2/(4 A_{z\perp})$, and $t_2^{opt} = \pi/(2 A_{z\parallel})$ (see Section S2 of the Supplementary Material for a full derivation). Starting from a state where the NV is in $m_S = +1$ and the nuclear spin is unpolarized, the system density matrix $\rho(t)$ before and after application of the CPMG train takes the form

$$\rho(t=0) = \frac{1}{4}(1 - P_0 + S_z) \to \rho(t_1^{opt}) = \frac{1}{4}(1 - P_0 + 2\alpha S_z I_\varphi) , \qquad (5)$$

where $P_0$ denotes the projection operator into $m_S = 0$ and $\alpha \equiv sgn(A_{z\perp})$. The last term in the expression for $\rho(t_1^{opt})$ corresponds to an antiphase nuclear spin coherence, i.e., a nuclear spin precession whose sign is conditioned on the orientation of the NV polarization. Importantly, the phase $\varphi$ of this coherence depends on the nuclear spin azimuthal coordinate, which, as we show below, will help us pinpoint the $^{13}$C location. To complete the polarization transfer, the system is allowed to evolve for a time $t_2^{opt}$ before applying an rf $\pi/2$-pulse. For the simpler case where $\phi = \varphi + m\pi$ with $m$ integer, the net result is the transformation

$$\rho(t_1^{opt}) \to \rho(t_2^{opt}) = \frac{1}{4}(1 - P_0)(1 + 2\delta I_z) , \qquad (6)$$

where we define $\delta \equiv sgn(A_{z\|}A_{z\perp}\cos(\varphi - \phi))$. The expression for $\rho(t_2^{opt})$ corresponds to a state where the NV is equally likely to point up or down (i.e., a depolarized state in the $m_S = \pm 1$ subspace) and the nuclear spin is polarized to $m_I = +1/2$ or $m_I = -1/2$ depending on the sign of $\delta$.

A numerical calculation of the nuclear spin response for an arbitrary number of pulses or duration of the free evolution interval is presented in Fig. 2a for the particular case of a $^{13}$C spin in a ~30 mT field ($\omega_I/2\pi = 320$ kHz) with spatial coordinates $r = 0.9$ nm, $\theta = 70.5°$, and $\varphi = 0$ (corresponding to $A_{z\perp}/2\pi = 26$ kHz and $A_{z\|}/2\pi = -19$ kHz). Complete nuclear spin polarization of alternating sign is only attained at multiples of $t_1^{opt}$ and $t_2^{opt}$, each of which reacting independently to a change of $A_{z\perp}$ or $A_{z\|}$ (see right panels in Fig. 2a). The latter is more clearly shown in Fig. 2b where we plot the calculated temporal coordinates of the first peak

transfer (in this example resulting in negative nuclear spin polarization) as we systematically vary either dipolar coupling constant. The analytical model agrees well with the numerical calculation in the (valid) regime where the nuclear Larmor frequency is sufficiently large (at least five times greater than the coupling). We also confirm the $2/\pi$ scaling of $A_{z\perp}$ anticipated in Eq. (3) using average Hamiltonian theory.

While in the calculations above mw pulses selectively act on the $m_S = \pm 1$ states, it is also possible to induce full nuclear polarization using a single quantum NV transition (i.e., $m_S = 0 \leftrightarrow m_S = \pm 1$), as we numerically demonstrate in Fig. 2c. However, rather than oscillating between states of positive and negative polarization, the nuclear spin polarizes or not depending on the chosen timing. Optimal transfer times comparable to those in Fig. 2a can be attained by increasing the dipolar coupling constants indicating that the down-scaling noted above is more pronounced for a single quantum transition. Further, both $t_1^{opt}$ and $t_2^{opt}$ are influenced by a change in either coupling constants (central and right panels in Fig. 2c) pointing to a more complicated relationship between the sequence timing and the strength of the dipolar interaction. The latter can be understood by noting that a single-quantum π-pulse is insufficient to change the sign of $S_z$ meaning that contributions of the form $AS_z I_z$ in Eq. (2) are not averaged out during the CPMG train.

We noted above that the nuclear/electron antiphase coherence generated by the CPMG train depends on the nuclear spin azimuthal angle $\varphi$ (see Eq. (5)). The latter has a direct impact on the resulting nuclear polarization, shown in Fig. 3a as we rotate the carbon location about the z-axis: For given evolution intervals $t_1$, $t_2$, the nuclear polarization exhibits a sinusoidal dependence on $\varphi$ whose amplitude ranges from the maximum possible (e.g., at the first peak transfer, red curve) to zero (e.g., at a nodal point, black curve). We interpret this behavior as a manifestation of the broken axial symmetry in the system Hamiltonian (Eqs. (1) and (2)), which, in turn, stems from the large difference between the electron and nuclear resonance frequencies. Fast manipulation of the paramagnetic center synchronic with $\omega_I$ during the CPMG train creates a resonant time-dependent field at the nuclear site, which leads to the appearance of nuclear spin (anti-phase) coherence. Since this field depends on the carbon azimuthal angle, it imprints the resulting nuclear spin coherence with a $\varphi$-dependent phase. Consequently, conversion to nuclear polarization requires control of $\phi$—the rf phase *in the laboratory frame*—easily attained in practice, e.g., by direct rf pulse synthesis. As we show below, this ability will prove key to identifying the spatial location of the nuclear spin.

Besides the rf phase, the interpulse separation in the CPMG train is another parameter worth commenting on. For the NV-$^{13}$C geometry considered above, Fig. 3b shows the dependence of the nuclear polarization $P_I$ on $\tau$ and $t_2$ for CPMG trains with different number $n$ of π-pulses. For $\omega_I \tau / 2\pi = 0.5$ and $\phi = 0$ (left panels), we find that the transfer efficiency is optimum at $n = 10$ and vanishes as $n \to 20$, in agreement with the results of Fig. 2a. However, nuclear polarization does build up as $\tau$ departs from half the $^{13}$C Larmor period. This effect is more pronounced for larger values of $n$ indicating that spin transfer to more weakly coupled nuclei is increasingly susceptible to imperfections in the timing of the CPMG train. This tendency is replicated in the case $\phi = 90$ (right panels), where the spin transfer is prevented only if $\tau$ precisely coincides with $\pi/\omega_I$.

To determine the nuclear spin coordinates we use a composite sequence where the protocol in Fig. 1 — designed to transfer spin polarization from the electron spin $S$ to the nuclear spin $I$ — is followed by NV re-initialization and nuclear spin polarization retrieval. The latter can be implemented in various ways, for example, by running the transfer protocol in reverse (Fig.

4a, see also Fig. S1 of the Supporting Material). By systematically increasing the polarization exchange intervals $t_1$ and $t_2$ and the rf phase $\phi$ one gathers a 3D data set that reflects on the efficiency of the transfer and hence on the coordinates of the target nuclear spin (Fig. 4b). Fourier transforming along $t_1$ and $t_2$ produces a set of extrema in the 3D space spanned by $A_{z\parallel}$, $A_{z\perp}$ and $\varphi$ (Fig. 4c), which can then be converted into real-space coordinates. The result is presented in Fig 4d for the example case of Figs. 2 and 3 (single spin with coordinates ($r = 0.9$ nm, $\theta = 70.5°$, $\varphi = 0°$)). Besides the 'real' nuclear site (dashed line in Fig. 4d), we identify an accompanying set of 'ghost' images, as expected in the case of a symmetric transfer/retrieval protocol (where the NV signal is insensitive to the absolute signs of $A_{z\parallel}$, $A_{z\perp}$ and $\phi$, Section S2.2 of the Supplementary Material); two possible sites result from each sign ambiguity, thus leading to a total of eight possible locations. For future reference, we also note the crescent shape of the isolevel surfaces, and the correspondingly poor azimuthal resolution.

All ambiguities on the true location of the nuclear site can be removed by breaking the symmetry between the transfer and retrieval segments of the protocol—a mirror image of each other in Fig. 4. For illustration purposes, Fig. 5a introduces an alternate retrieval sequence where the cyclic application of inversion pulses makes the effective Hamiltonian equal to $H_1^* = \frac{2}{\pi} A_{z\perp} S_z I_\varphi$ during both evolution intervals, $\tilde{t}_1$ and $\tilde{t}_2$. Unlike the result in Figs. 4c and 4d, it can be shown that the accompanying NV signal has an overall sign that depends on $\alpha \equiv sgn\{A_{z\perp}\}$, and $\beta \equiv sgn\{A_{z\parallel}\}$ but not on $\gamma \equiv sgn\{\cos(\phi - \phi)\}$. Other combinations are possible with complementary pulse protocols, which allows one to independently determine $\alpha$, $\beta$, and $\gamma$ (and correspondingly the 'true' nuclear site). The result is presented in Fig. 5b for the more complex case of multiple $^{13}$C spins weakly coupled to an NV (Section S2 of the Supplementary Material).

An important practical consideration in the application of this technique is the limit spatial resolution, in general, a complex function of various parameters. Possibly the most relevant factor is the NV lifetime, of immediate impact on the duration of the NV response and thus on the broadening of the resulting 'peaks' in the 3D spectrum. An example is presented in Fig. 5c, where we compare the 'images' of the same $^{13}$C cluster for varying NV lifetimes. Conversely, for a fixed NV lifetime, the spatial resolution decreases as the hyperfine coupling weakens, simply because a longer interaction time is required to exchange polarization with the target nuclear spin. For example, additional calculations (not shown here for brevity) indicate that the volume enclosed by the isolevel surface quadruples when increasing the nuclear distance $r$ by a factor ~1.8, from 0.9 nm to 1.6 nm. Naturally, similar effects are observed as the polar angle approaches $\theta = 0°$ or $\theta = 54.7°$, respectively the nodes of $A_{z\perp}$ and $A_{z\parallel}$.

Further work will be needed to circumvent some present limitations, including the extended crescent shape—and correspondingly moderate resolution—along the azimuthal angle $\varphi$ (Fig. 4d). This coordinate differs from the other two in that it is defined by the phase selectivity of the nuclear spin projection pulses throughout the polarization exchange ($P_S \propto \cos^2(\varphi - \phi)$, Section S2.2 of the Supplementary Material). Higher azimuthal selectivity may be attained, for instance, via the use of composite rf π/2-pulses more sensitive to the nuclear spin phase.

The ideas underlying the technique presented herein can be extended in several complementary directions. First, we emphasize that spins other than the NV can be envisioned as the source of nuclear spin polarization. Besides the various paramagnetic centers mentioned above, one possibility is the use of 'dark' electronic spins such as the P1 center in diamond (formed by a substitutional nitrogen), already shown to polarize by contact with the NV[21].

Alternatively, one can envision the use of paramagnetic labels (e.g., nitroxyls) to mediate the interaction between the NV and nuclear spins in outer molecules, as demonstrated recently[16]. Along these lines, we mention that our technique could be conceivably adapted to determine the spatial positions of electronic rather than nuclear spins. Such strategy would be helpful in pinpointing the positions of key sites within molecules tethered to the diamond surface[22]. Further, since the polarization transfer efficiency depends on the exact spatial coordinates of the target (electronic or nuclear) spin, introducing a variable time interval between polarization transfer and retrieval could be exploited to monitor various dynamical processes. Examples are slow molecular folding processes activated externally either by chemical or optical means.

Purely as a polarization transfer strategy, our approach could serve as a route to dynamically polarize organic systems adsorbed on the diamond surface[23]. In particular, because the pulse timing can be controlled with nanosecond precision (much shorter than the microsecond nuclear Larmor period at moderate magnetic fields), the time jitter is expected to be low, thus ensuring high spin transfer efficiency. The latter may prove an advantage when compared to, e.g., the Hartman-Hahn scheme[24], where amplitude fluctuations of the rf field reduce the probability of a flip-flop, especially for weakly coupled nuclei. The flipside, however, is the greater sensitivity to inter-nuclear spin interactions, which, in the present protocol, must remain smaller than the coupling to the NV. While this is typically the case within the diamond crystal (where $^{13}$C spins are dilute), the opposite limit applies to nuclei on the diamond surface (such as $^{1}$H spins in adsorbed molecules). This complication could be somewhat mitigated, e.g., via the combined use of inter-nuclear dynamical decoupling sequences.

The authors acknowledge support from the National Science Foundation through grants NSF-1314205 and NSF-1309640.

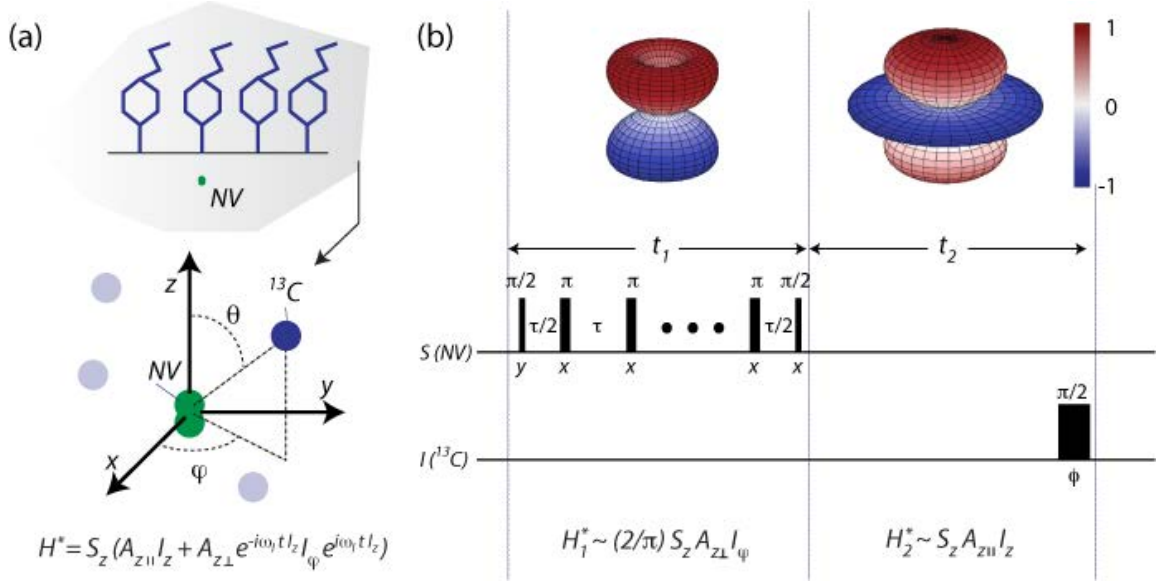

**Fig. 1: Spatial encoding via spin polarization transfer.** (a) Schematics of an NV center and interacting nuclear spins. One intriguing possibility is the use of shallow NVs to locate spin-labeled nuclei from molecules on the diamond surface. $H^*$ denotes the double-rotating-frame Hamiltonian governing an individual NV-$^{13}$C pair interacting via a weak hyperfine coupling. (b) Double resonance pulse protocol. A magnetic field $B$ is applied along $z$, coincident with the NV symmetry axis. The NV is initialized into $m_S = 1$ and all microwave pulses induce transitions between the $m_S = \pm 1$ states. $\rho$ denotes the NV-$^{13}$C density matrix, here evaluated at different points of the pulse protocol, and $H_i^*$, $i = 1,2$, is the effective time-independent Hamiltonian during each half of the pulse sequence. In each case, the top graph indicates the normalized $^{13}$C dipolar coupling as a function of the azimuthal and polar angles (respectively, $\varphi$ and $\theta$ in (a)).

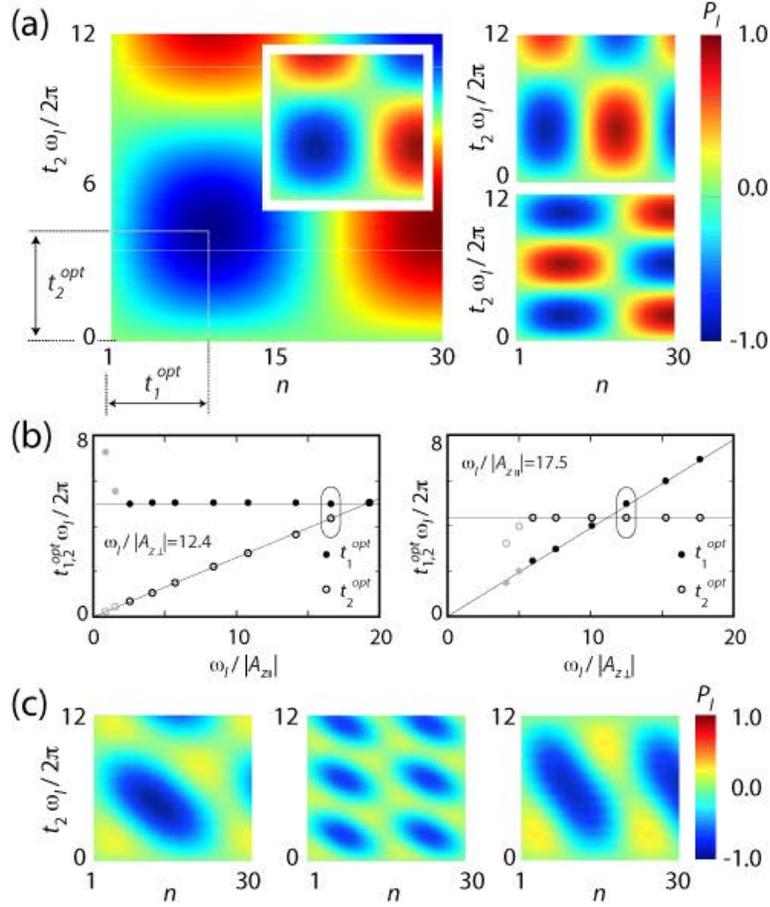

**Fig. 2: Pulsed spin polarization transfer to $^{13}$C.** (a) (Right) Numerically calculated $^{13}$C spin polarization after application of the pulse sequence in Fig. 1. The initial NV state is $m_S$=1, the applied magnetic field is 30 mT, and the hyperfine coupling constants are $A_{z\perp}/2\pi = 25.9$ kHz, $A_{z\parallel}/2\pi = -18.3$ kHz corresponding to spatial coordinates $(r = 0.9$ nm, $\theta = 70.5°$, $\varphi = 0°)$. The insert represents the nuclear polarization as derived from average Hamiltonian theory (Eq. (S15) in Section S2 of the supplementary material). (Left) Numerically calculated $^{13}$C spin polarization assuming $A_{z\perp}/2\pi = 39.0$ kHz, $A_{z\parallel}/2\pi = -18.3$ kHz (top panel) and $A_{z\perp}/2\pi = 25.9$ kHz, $A_{z\parallel}/2\pi = -38.0$ kHz (lower panel). (b) Optimum transfer times $t_1^{opt}$ and $t_2^{opt}$ (full and empty circles, respectively) as a function of $\omega_I/A_{z\parallel}$ (right plot, $A_{z\perp}/2\pi = 25.9$ kHz) and $\omega_I/A_{z\perp}$ (left plot, $A_{z\parallel}/2\pi = -18.3$ kHz). The circled pair of dots corresponds to the main case in (a). Lines indicate the relations $t_1^{opt} = \pi^2/(4A_{z\perp})$ and $t_2^{opt} = \pi/(2A_{z\parallel})$ serving as a guide to the eye. (c) Same as in (a) but for mw pulses acting on the NV $m_S = 0 \leftrightarrow m_S = +1$ transition. The $^{13}$C angular coordinates are the same but the distance to the NV is 0.7 nm corresponding to $A_{z\perp}/2\pi = 55$ kHz, $A_{z\parallel}/2\pi = -39$ kHz. In (a) through (c) the rf-phase is $\phi = 0$, and the separation between π-pulses is $\tau = \pi/\omega_I$.

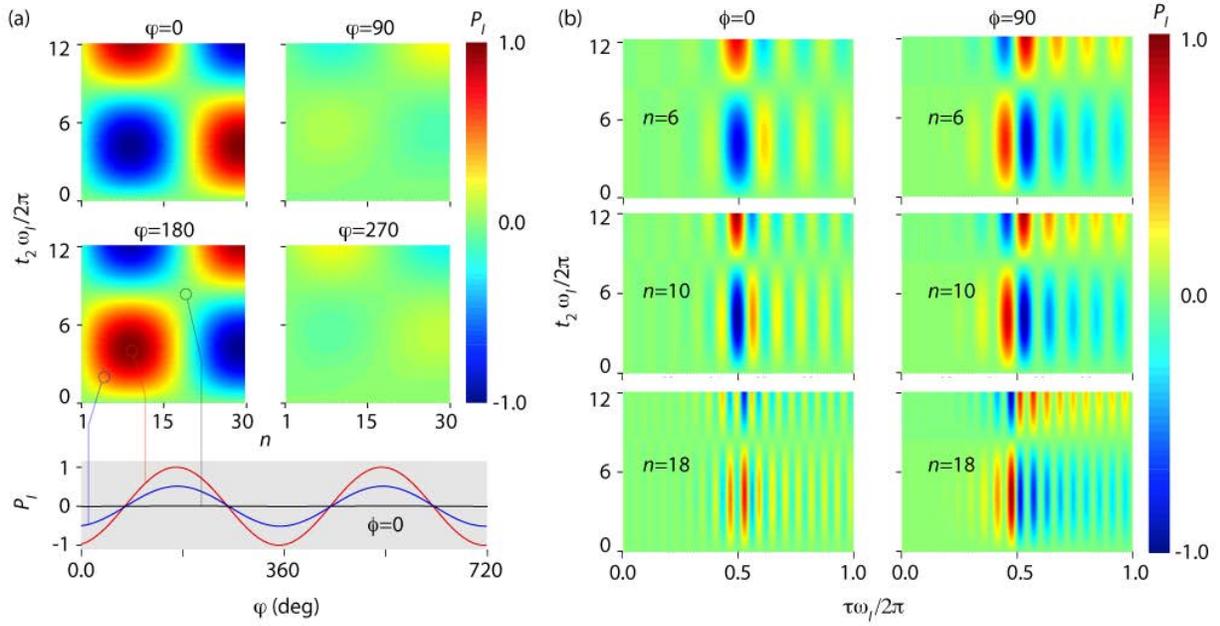

**Fig. 3: The role of rf phase and mw timing.** (a) Upper panels: Nuclear spin polarization $P_I$ as a function of $n$, the number of π-pulses, and $t_2$ for a $^{13}$C spin with variable azimuthal coordinate $\varphi$. The rf phase is $\phi = 0$. A similar dependence is obtained for a fixed nuclear spin site as one changes the rf phase. Lower panel: $P_I$ as a function of $\varphi$ for $n = 10$, $\omega_I t_2/2\pi = 4.2$ (red curve), for $n = 6$, $\omega_I t_2/2\pi = 2.1$ (blue curve), and for $n = 19$, $\omega_I t_2/2\pi = 8.2$ (black curve). In all cases the rf phase is $\phi = 0$. (b) Nuclear spin polarization as a function of the normalized inter-pulse separation $\tau$ and free evolution interval $t_2$ for CPMG trains with $n$ π-pulses. Plots on the left (right) column correspond to rf phase $\phi = 0$ ($\phi = 90$). In (a) and (b) the NV interacts with $^{13}$C spin located at ($r = 0.9$ nm, $\theta = 70.5°$, $\varphi = 0°$).

Laraoui et al., Fig. 4

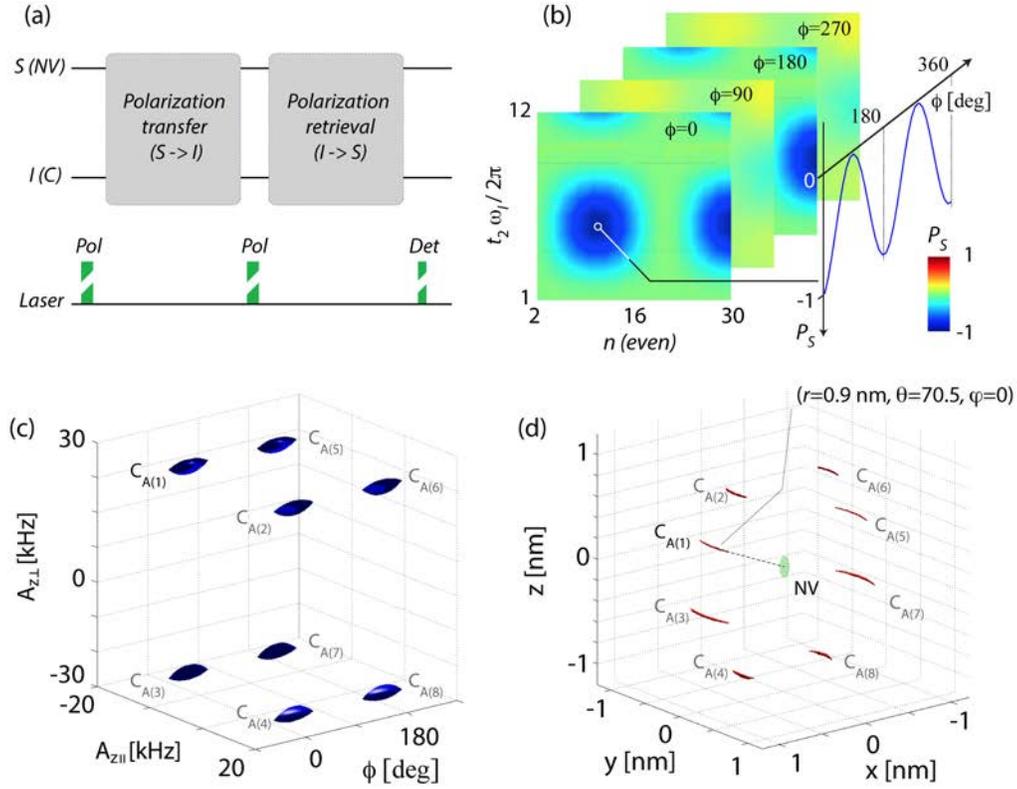

**Fig. 4: Determining the nuclear spin position.** (a) Schematics of the pulse sequence. Discontinued laser pulses indicate NV polarization into (or detection of) $m_S=+1$. (b) NV response $P_S$ as a function of the polarization exchange intervals $t_1$ and $t_2$ and rf phase $\phi$ for a single $^{13}$C spins with coordinates ($r = 0.9$ nm, $\theta = 70.5°$, $\varphi = 0°$). For clarity, only one maximum is shown but we assume a full data set whose decay is governed by the NV lifetime. (c) Iso-level plot of the 3D spectrum (magnitude mode) upon Fourier transform of the data set in (a) along the time dimensions $t_1$ and $t_2$. The corresponding 3D map with the $^{13}$C location (right) can be calculated via the transformation $(A_{z\|}, A_{z\perp}, \phi) \rightarrow (r, \theta, \varphi)$. Besides the 'real' position (labeled $C_{A(1)}$ in the 3D plots) there are seven 'ghost' images ($C_{A(2)}\ldots C_{A(8)}$) arising from a sign ambiguity in each dimension. All surfaces in (c) and (d) correspond to 80% of the NV signal maximum.

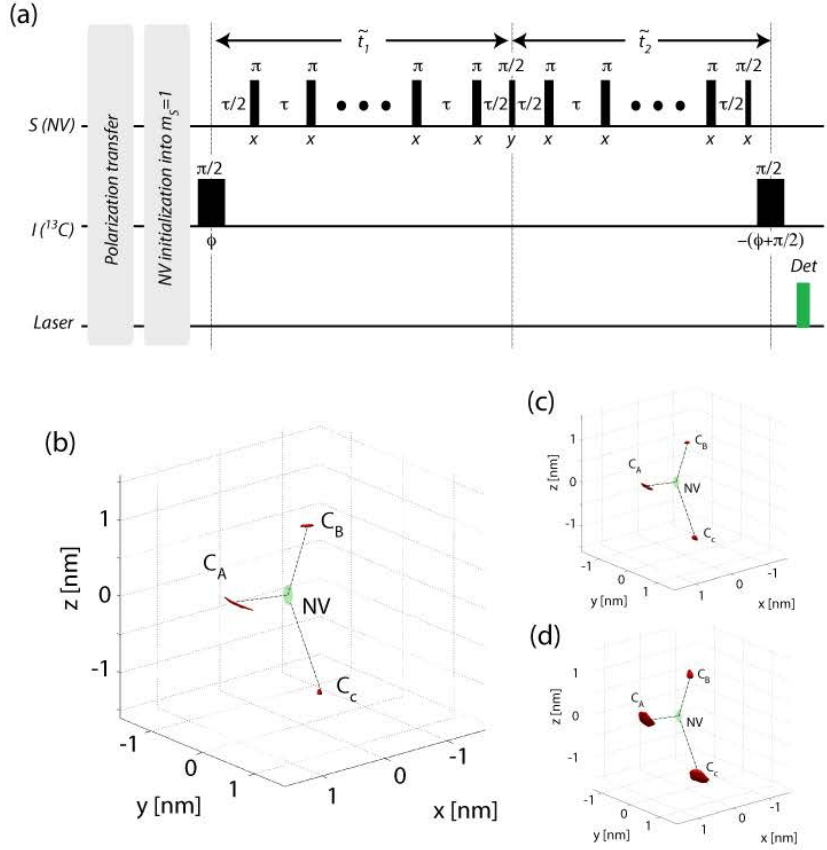

**Fig. 5:** (a) All ambiguities in the nuclear spin location can be removed with the aid of alternate polarization retrieval protocols. The schematics shows one such protocol where the NV signal is imprinted with the relative sign of the hyperfine coupling constants (see text and Section S2 of the supplementary material). (b) We determine the locations of three $^{13}$C spins (denoted as $C_A$, $C_B$, $C_C$) weakly coupled to an NV (represented by a central green dot). Using three complementary retrieval protocols (including that shown in (a)) we eliminate all ghost images. As in Fig. 4, the red surfaces correspond to isolevel plots of the transformed spectrum at 80% of the NV signal maximum. (c-d) Same as in (b) but for NV coherence lifetimes of 37 µs and 19 µs, respectively.

# Imaging nuclear spins weakly coupled to a probe paramagnetic center

*Abdelghani Laraoui[1], Daniela Pagliero[1], Carlos A. Meriles[1]*

[1]Dept. of Physics, CUNY-City College of New York, New York, NY 10031, USA.

## S1. Determining the effective Hamiltonian

The spin dynamics of an NV center coupled to a distant $^{13}$C nucleus can be described via the Hamiltonian

$$H = \Delta S_z^2 - \gamma_S B_0 S_z - \gamma_I B_0 I_z + H_d, \qquad (S1)$$

where we follow the notation in Eq. (1) of the main text, and $H_d$ represents the dipolar interaction between the NV and the $^{13}$C spin (we assume the hyperfine interaction sufficiently weak so as to ignore all contact contributions). In the typical regime where the combined Zeeman and crystal field dominates, i.e., when $|\omega_S| \gg \|H_d\|$, we can truncate all terms non-diagonal on $S_z$ (the secular approximation). Eq. (S1) can then be recast as

$$H = \Delta S_z^2 - \gamma_S B_0 S_z - \gamma_I B_0 I_z + A_{z\parallel} S_z I_z + A_{z\perp} S_z I_\varphi, \qquad (S2)$$

as presented in the main text. Of note, the secular approximation is not applicable to the nuclear spin because, as we show next, NV manipulation synchronous with the nuclear Larmor frequency can re-introduce off-diagonal terms into the nuclear spin dynamics. Besides coupling to the $^{13}$C spin, the NV spin is coupled to the nuclear spin of the $^{14}$N (or $^{15}$N) host. This hyperfine coupling has the isotropic form $\tilde{A} S_z K_z$, where $K_z$ denotes the z-projection of the nitrogen nuclear spin operator. Though the amplitude $\tilde{A}$ of this interaction (~2-3 MHz, depending on the nitrogen isotope) is much stronger than the class of NV-$^{13}$C couplings considered herein ($A_{z\parallel}, A_{z\perp} \sim 50$ kHz or less), the nitrogen spin can be considered pinned in an eigenstate throughout the duration of the protocol ($\tilde{A} S_z K_z$ commutes with all terms in Eq. (S2)). Correspondingly, the hyperfine coupling with the nitrogen nucleus leads to an effective magnetic field $\delta B_0 = \tilde{A} K_z$, which can be ignored when the microwave field $B_1$ governing the NV evolution is sufficiently strong (~0.3 mT or greater, the case in most experiments). Of note, the nitrogen nuclear spin can be initialized into a pre-defined target state[1], in which case $\delta B_0$ takes a fixed, known value, which can be seamlessly assimilated into the main Zeeman field.

In the doubly rotating frame resonant with the carbon Larmor frequency and the chosen NV transition (e.g., $|m_S = 1\rangle \leftrightarrow |m_S = -1\rangle$ or $|m_S = 0\rangle \leftrightarrow |m_S = -1\rangle$), Eq. (S2) takes the time-dependent form

$$H^*(t) = A_{z\parallel} S_z I_z + A_{z\perp} S_z e^{-i\omega_I t I_z} I_\varphi e^{i\omega_I t I_z} , \quad (S3)$$

where we follow the notation in Eqs. (1) and (2) of the main text. The formal solution for the evolution operator is given by

$$U(t, t_0) = T\exp\left(-i \int_{t_0}^{t} dt' H^*(t')\right) , \quad (S4)$$

where $T$ denotes the time-ordering operator. To determine the system dynamics during application of the polarization transfer protocol (Fig. 1), let's first consider the system evolution during $t_1$, throughout the application of the CPMG train. During one unit of the train $[\tau/2 - R_S(\pi) - \tau - R_S(\pi) - \tau/2]$, the evolution operator takes the form

$$U(t_0 + 2\tau, t_0) = U\left(t_0 + 2\tau, t_0 + \frac{3}{2}\tau\right) R_S(\pi) U\left(t_0 + \frac{3}{2}\tau, t_0 + \frac{1}{2}\tau\right) R_S(\pi) U\left(t_0 + \frac{1}{2}\tau, t_0\right), \quad (S5)$$

where $R_S(\pi)$ denotes a $\pi$-pulse acting on the chosen NV transition. Here we assume the condition $\pi/\tau \sim \omega_I \gg A_{z\parallel}, A_{z\perp}$, allowing us to use the Magnus expansion to find an approximate expression. For example, considering the interval $a = \left(t_0, t_0 + \frac{1}{2}\tau\right)$, we have

$$U\left(t_0 + \frac{1}{2}\tau, t_0\right) = \exp\left(-i\frac{\tau}{2}\left(H_a^{*(0)} + H_a^{*(1)} + H_a^{*(2)} + \cdots\right)\right) \approx \exp\left(-i\frac{\tau}{2} H_a^{*(0)}\right) , \quad (S6)$$

where

$$H_a^{*(0)} = \frac{2}{\tau} \int_{t_0}^{t_0 + \frac{\tau}{2}} dt' H^*(t') = \quad (S7)$$

$$= A_{z\parallel} S_z I_z + \frac{2 A_{z\perp} S_z}{\tau \omega_I} \left\{ I_\varphi \left(\sin \omega_I \left(t_0 + \frac{\tau}{2}\right) - \sin(\omega_I t_0)\right) + I_{\varphi + \pi/2} \left(\cos \omega_I \left(t_0 + \frac{\tau}{2}\right) - \cos(\omega_I t_0)\right) \right\} ;$$

and we have ignored the higher order terms since $\left\|H_a^{*(j)}\right\| \sim \left(\frac{A}{\omega_I}\right)^j \left\|H_a^{*(0)}\right\| \ll \left\|H_a^{*(0)}\right\|$ for $j = 1,2,...$ and $A = max\{A_{z\|}, A_{z\perp}\}$. By the same token, during intervals $b = \left(t_0 + \frac{1}{2}\tau, t_0 + \frac{3}{2}\tau\right)$ and $c = \left(t_0 + \frac{3}{2}\tau, t_0 + 2\tau\right)$, the average Hamiltonians are given by

$$H_b^{*(0)} = A_{z\|}S_zI_z + \frac{A_{z\perp}S_z}{\tau\omega_I}\left\{I_\varphi\left(\sin\omega_I\left(t_0 + \frac{3\tau}{2}\right) - \sin\omega_I\left(t_0 + \frac{\tau}{2}\right)\right) + \right.$$

$$\left. +I_{\varphi+\pi/2}\left(\cos\omega_I\left(t_0 + \frac{3\tau}{2}\right) - \cos\omega_I\left(t_0 + \frac{\tau}{2}\right)\right)\right\}, \quad (S8)$$

and

$$H_c^{*(0)} = A_{z\|}S_zI_z + \frac{2A_{z\perp}S_z}{\tau\omega_I}\left\{I_\varphi\left(\sin\omega_I(t_0 + 2\tau) - \sin\omega_I\left(t_0 + \frac{3\tau}{2}\right)\right) + \right.$$

$$\left. +I_{\varphi+\pi/2}\left(\cos\omega_I(t_0 + 2\tau) - \cos\omega_I\left(t_0 + \frac{3\tau}{2}\right)\right)\right\}. \quad (S9)$$

Therefore, during one CPMG unit the average Hamiltonian is given by

$$H_1^* \approx \frac{1}{4}H_a^{*(0)} + R_S(\pi)H_b^{*(0)}R_S^\dagger(\pi) + \frac{1}{4}H_c^{*(0)}. \quad (S10)$$

Assuming all mw pulses act on the $|m_S = 1\rangle \leftrightarrow |m_S = -1\rangle$ transition (see main text for references), we have $S_z \xrightarrow{R_S(\pi)} -S_z$, and Eq. (S10) takes the form

$$H_1^* = \frac{A_{z\perp}S_z}{2\tau\omega_I}\left\{I_\varphi\left(2\sin\omega_I\left(t_0 + \frac{\tau}{2}\right) - 2\sin\omega_I\left(t_0 + \frac{3\tau}{2}\right) + \sin\omega_I(t_0 + 2\tau) - \sin(\omega_I t_0)\right) + \right.$$

$$\left. +I_{\varphi+\pi/2}\left(2\cos\omega_I\left(t_0 + \frac{\tau}{2}\right) - 2\cos\omega_I\left(t_0 + \frac{3\tau}{2}\right) + \cos\omega_I(t_0 + 2\tau) - \cos(\omega_I t_0)\right)\right\}. \quad (S11)$$

Typical conditions are $\omega_I t_0 = 2m\pi$ with $m$ integer and $\omega_I \tau = \pi$ thus leading to

$$H_1^* = \frac{2}{\pi}A_{z\perp}S_zI_\varphi. \quad (S12)$$

On the other hand, a similar average Hamiltonian calculation during during $t_2$ leads to the effective Hamiltonian

$$H_2^* = A_{z\parallel} S_z I_z + \frac{A_{z\perp} S_z}{t_2 \omega_I} \{I_\varphi (\sin \omega_I (t_2 + t_0) - \sin \omega_I t_0) +$$

$$+ I_{\varphi + \pi/2} ((\cos \omega_I (t_2 + t_0) - \cos \omega_I t_0))\} \approx A_{z\parallel} S_z I_z. \quad \text{(S13)}$$

## S2. Dynamics of spin polarization transfer and retrieval

*i-Polarization transfer*

Assuming that the NV is initialized into the $|m_S = +1\rangle$ and that the nuclear spin is in a fully mixed state, we describe the initial state of an NV-$^{13}$C pair via the density matrix

$$\rho_0 = \frac{1}{4} (1 - P_0 + S_z), \quad \text{(S14)}$$

where $P_0$ denotes the projection operator into the $|m_S = 0\rangle$ state. The system evolution during the first half of the polarization transfer protocol (Fig. 1 of the main text) can then be described via the successive transformations

$$\rho_0 \xrightarrow{(\pi/2)_y^S} \frac{1}{4}(1 - P_0 + S_x) \xrightarrow{A_1 S_z I_\varphi} \frac{1}{4}(1 - P_0 + S_x \cos A_1 t_1 + 2 I_\varphi S_y \sin A_1 t_1) \xrightarrow{(\pi/2)_x^S}$$

$$\rho_1 = \frac{1}{4}(1 - P_0 + S_x \cos A_1 t_1 + 2 I_\varphi S_z \sin A_1 t_1), \quad \text{(S15)}$$

where we define $A_1 \equiv \frac{2}{\pi} A_{z\perp}$. By the same token, the system evolution during $t_2$ is given by

$$\rho_1 \xrightarrow{A_2 S_z I_z} \frac{1}{4}\left(1 - P_0 + \cos A_1 t_1 \left(S_x \cos A_2 t_2 + 2 I_z S_y \sin A_2 t_2\right) + \right.$$

$$+ 2 \sin A_1 t_1 \left(I_\varphi S_z \cos A_2 t_2 + I_{\varphi + \pi/2} S_z^2 \sin A_2 t_2\right)\Big), \quad \text{(S16)}$$

where $A_2 \equiv A_{z\parallel}$. Using $\phi$ to denote the phase of the rf $\pi/2$-pulse and noting that $I_\varphi = I_\phi \cos(\varphi - \phi) + I_{\phi + \pi/2} \sin(\varphi - \phi)$ and $I_{\phi + \pi/2} = -I_\phi \sin(\varphi - \phi) + I_{\phi + \pi/2} \cos(\varphi - \phi)$, the density matrix at the end of the polarization transfer is

$$\rho_2 = \frac{1}{4}\Big(1 - P_0 + S_x \cos A_1 t_1 \cos A_2 t_2 - 2 I_{\phi + \pi/2} S_y \cos A_1 t_1 \sin A_2 t_2 +$$

$$+ 2 S_z \sin A_1 t_1 \cos A_2 t_2 \left(I_\phi \cos(\varphi - \phi) + I_z \sin(\varphi - \phi)\right) +$$

$$+ 2 S_z^2 \sin A_1 t_1 \sin A_2 t_2 \left(-I_\phi \sin(\varphi - \phi) + I_z \cos(\varphi - \phi)\right)\Big). \quad \text{(S17)}$$

When $|A_1 t_1| = |A_2 t_2| = \pi/2$ and $\phi = \varphi + m\pi$, Eq. (S17) simplifies to

$$\rho_2^{opt} = \frac{1}{4}(1 - P_0)(1 + 2\delta I_z),  \qquad (S18)$$

where $\delta = sgn\{A_{z\perp} A_{z\parallel} \cos m\pi\}$, $m$ is an integer, and the subscript indicates optimal polarization transfer to the nuclear spin. An analytical expression for the resulting nuclear spin polarization can be calculated from Eq. (S17) using the relation $P_I = 2\, tr\{\rho_2 I_z\}$; one obtains

$$P_I = \sin A_1 t_1 \sin A_2 t_2 \cos(\varphi - \phi). \qquad (S19)$$

*ii-Polarization retrieval*

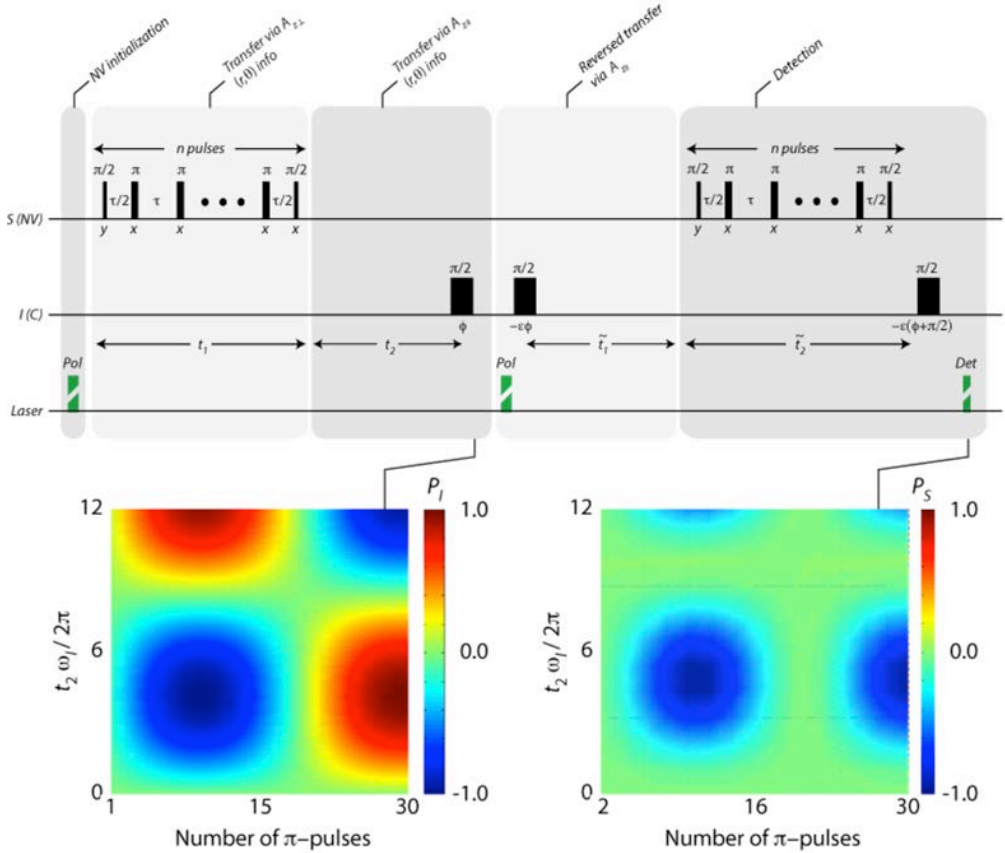

**Fig. S1:** Schematics of the polarization transfer and retrieval protocol used in Fig. 4 of the main text. The lower graphs indicate the nuclear and NV spin polarization ($P_I$ and $P_{NV}$, respectively) at the end of the transfer and retrieval segments of the sequence for the case $\varphi = \phi = 0$ and $\varepsilon = -1$. Discontinued laser pulses indicate NV polarization into (or detection of) $m_S = +1$.

To determine the spatial position of the nuclear spin, we use a composite protocol where the spin polarization transferred to the nucleus during the first half is sent back to the NV during the second half for inspection. One version of the full pulse sequence is presented in Fig. S1 where, the nuclear spin polarization is probed by re-initializing the NV and running the transfer protocol in reverse. To derive analytical expressions for the resulting signal, we model the NV optical pumping after the transfer by tracing the density matrix operator over the NV states and subsequently re-setting the NV into $\pm|m_S = 1\rangle$; the latter corresponds to a non-unitary operation that emulates NV relaxation in the presence of optical illumination. For simplicity, we assume that the light pulse destroys the nuclear spin coherence without, however, affecting the nuclear spin polarization (consistent with the weak NV-$^{13}$C coupling assumed throughout the text). Under these conditions, Eq. (S17) takes the form

$$\tilde{\rho}_0 = \frac{1}{4}(1 - P_0 + \eta S_z)(1 + 2n_1 n_2 c_\varphi I_z), \qquad (S20)$$

where $\eta = \pm 1$ indicates NV initialization into $\pm|m_S = 1\rangle$, and we use the simplified notation $n_j \equiv \sin A_j t_j$, $j = 1,2$ and $c_\varphi \equiv \cos(\varphi - \phi)$ (subsequently we denote $c_j \equiv \cos A_j t_j$, $j = 1,2$ and $n_\varphi \equiv \sin(\varphi - \phi)$). Starting from Eq. (S20), we carry out a calculation similar to that above (Section S2.i) and derive the system density matrix at the end of the retrieval sequence $\tilde{\rho}_2$. The resulting NV signal is then given by

$$P_S = tr\{\tilde{\rho}_2 S_z\} = \epsilon(n_1 n_2 c_\varphi)^2 - \eta \epsilon n_1^2 n_2 c_2 c_\varphi n_\varphi, \qquad (S21)$$

where the above expression is valid for an even number of π-pulses and we assume for simplicity $\tilde{t}_1 = t_2$ and $\tilde{t}_2 = t_1$ (compare with Fig. 4 in the main text). Note that the second term can be cancelled by cycling the NV initialization from $|m_S = +1\rangle$ to $|m_S = -1\rangle$ (corresponding to changing $\eta$ in Eq. (S21) from $\eta = 1$ to $\eta = -1$), which leads to the simplified expression

$$P_S = \epsilon(n_1 n_2 c_\varphi)^2. \qquad (S22)$$

Altering the rf phase from $\varepsilon = 1$ to $\varepsilon = -1$ introduces an overall sign change in the NV signal,

which, experimentally, can be used to separate the nuclear spin contribution from any undesired background.

### iii-Spatial localization of the nuclear spin

Eq. (S22) is insensitive to the absolute signs of the hyperfine constants $A_{z\parallel}, A_{z\perp}$ and cannot distinguish between nuclei whose azimuthal coordinates $\varphi$ differ by 180 degrees. The latter leads to eight possible positions for the nuclear spin, thus preventing the unequivocal determination of the nuclear coordinates. To remove all ambiguities, we implement three auxiliar pulse protocols, summarized in Fig. S2. For illustration purposes, we start by considering the first protocol in the figure (which we refer to as AP1). After polarization transfer and NV reinitialization, the system density matrix is given by Eq. (S20). During the modified polarization retrieval, the system evolution can be followed via the set of transformations

$$\tilde{\rho}_0 \xrightarrow{(\pi/2)^I_{\varepsilon\phi}} \frac{1}{4}(1 - P_0 + S_z)(1 - 2\alpha\beta\varepsilon I_{\varphi+\pi/2}) \xrightarrow{A_1 S_z I_\varphi} \frac{1}{4}(1 - P_0 + S_z)(1 + 2\beta\varepsilon I_\varphi S_z)$$

$$\xrightarrow{(\pi/2)^S_y} \xrightarrow{A_1 S_z I_\varphi} \xrightarrow{(\pi/2)^S_x} \frac{1}{4}(1 - P_0 + \alpha\beta\varepsilon S_z)(1 + 2\beta\varepsilon I_\varphi)$$

$$\xrightarrow{(\pi/2)^I_{-\varepsilon(\phi+\pi/2)}} \tilde{\rho}_2^{AP1} = \frac{1}{4}(1 - P_0 + \alpha\beta\varepsilon S_z)(1 + 2\beta\gamma I_z) \,, \qquad (S23)$$

where, for simplicity, we have restricted the calculation to the situation where the spin exchange is optimal, namely $|A_1 \tilde{t}_1| = |A_2 \tilde{t}_2| = \pi/2$ and $\phi = \varphi + m\pi$, and we use the notation $\alpha \equiv sgn\{A_{z\perp}\}$, $\beta = sgn\{A_{z\parallel}\}$, and $\gamma = sgn\{\cos(\varphi - \phi)\}$. Note that in this case the NV signal $P_S^{AP1} = tr\{\tilde{\rho}_2^{AP1} S_z\} = \alpha\beta\varepsilon$ is imprinted by the relative sign of both hyperfine coupling constants. Similar calculations can be carried out for the alternate protocols AP2 and AP3. We obtain

$$\tilde{\rho}_0 \xrightarrow{AP2} \tilde{\rho}_2^{AP2} = \frac{1}{4}(1 - P_0 + \alpha\beta\gamma S_z)(1 + 2\beta\varepsilon I_z) \,, \qquad (S24)$$

and

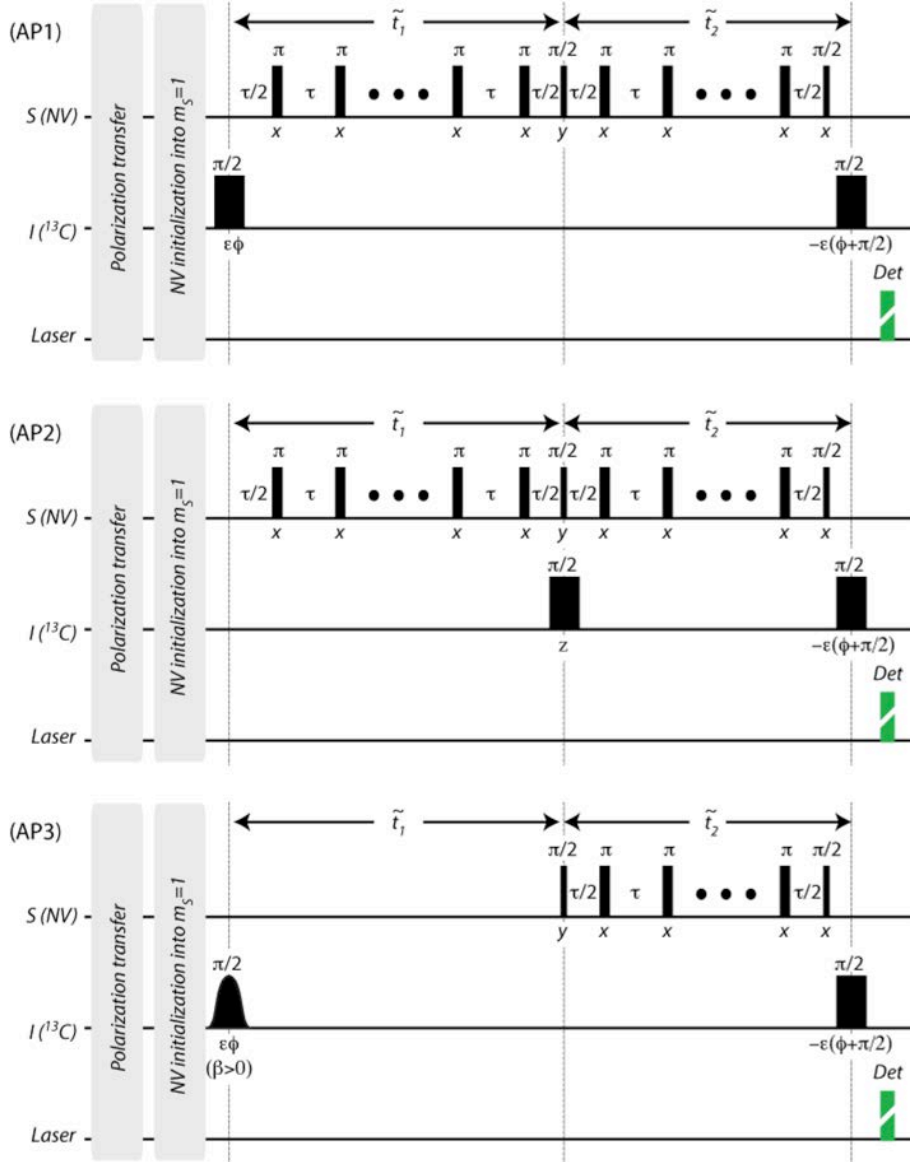

**Fig. S2**: Complementary spin polarization retrieval protocols. Upon polarization transfer to the nuclear spin and NV re-initialization, alternate retrieval schemes imprint the NV signal with a sign that depends on the hyperfine constants and the azimuthal angle. All three coordinates of the nuclear spin site can be determined with no ambiguity by comparing the results from these three protocols. Discontinued laser pulses indicate NV detection of $m_S=+1$.

$$\tilde{\rho}_0 \xrightarrow{AP3} \tilde{\rho}_2^{AP3} = \begin{cases} \frac{1}{4}(1 - P_0 + \varepsilon S_z)(1 + 2\alpha\gamma I_z) \; if \; \beta > 0 \\ \frac{1}{4}(1 - P_0)(1 + 4\beta\varepsilon I_z I_{\varphi+\pi/2}) \; if \; \beta < 0 \end{cases}, \quad (S25)$$

which leads to NV signals $P_S^{AP2} = \alpha\beta\gamma$, and $P_S^{AP3} = \varepsilon$ if $\beta > 0$ or $P_S^{AP3} = 0$ if $\beta < 0$. Therefore, with $\beta$ determined from AP3, we find the sign $\alpha$ from AP1 and finally $\gamma$ from the result in AP2.

*iv-Interaction with multiple nuclear spins*

To illustrate how the presence of more than one nuclear spin alters the system dynamics we consider the case of an NV interacting with two weakly coupled $^{13}$C atoms. In this case the effective Hamiltonians during each half of the polarization transfer (Fig. 1, main text), are given by

$$H_1^* = \frac{2}{\pi}\left(A_{z\perp} S_z I_\varphi + A'_{z\perp} S_z I'_\varphi\right), \quad (S26)$$

and

$$H_2^* = A_{z\|} S_z I_\varphi + A'_{z\|} S_z I'_\varphi, \quad (S27)$$

where we use a prime to distinguish operators associated with the second nuclear spin, and we follow the same notation as in Eqs. (S12) and (S13). Similar to Eqs. (S17) and (S20) we find the density matrix after the spin transfer and NV re-initialization

$$\tilde{\rho}_0 = \frac{1}{8}(1 - P_0 + \eta S_z)\left(1 + 2c'_1 n_1 n_2 c_\varphi I_z + 2c_1 n'_1 n'_2 c'_\varphi I'_z\right). \quad (S28)$$

As before, we assume that the nuclear spin coherences are destroyed during the NV optical pumping, and use a shortened notation for the trigonometrical functions (see Eq. (S20)). After some algebra, we find the density matrix after polarization retrieval

$$\tilde{\rho}_2 = \frac{\varepsilon}{8}\left(\left(c'_1 n_1 n_2 c_\varphi\right)^2 S_z + \left(c_1 n'_1 n'_2 c'_\varphi\right)^2 S_z + n.c.\right), \quad (S29)$$

where, as before, we assume $\tilde{t}_1 = t_2$, $\tilde{t}_2 = t_1$, and *n.c.* stands for terms not contributing to the final NV signal (which can now be formally calculated via the relation $P_S = tr\{\tilde{\rho}_2 S_z\}$).

Comparison of Eqs. (S22) and (S29) indicates tthat the polarization transfer to a given target carbon spin is less than optimum because of the simultaneous interaction of the NV with the other carbons in the bath. The latter leads to a reduction of the signal contrast as the number of nuclear spins increases. This problem may be circumvented by making the polarization transfer $^{13}$C-selective, e.g., by applying an rf π-pulse at the midpoint of $t_1$ to selectively invert all nuclear spins except the target $^{13}$C spin. An alternate strategy—valid for mw pulses acting on the $|m_S = 0\rangle \leftrightarrow |m_S = +1\rangle$ (or $|m_S = 0\rangle \leftrightarrow |m_S = -1\rangle$) transition—could be to properly choose the time interval $\tau$ between π-pulses in the CPMG train.

---

[1] D. Pagliero, A. Laraoui, J. Henshaw, C.A. Meriles, "Recursive polarization of nuclear spins in diamond at arbitrary magnetic fields", *Appl. Phys. Lett.* **105**, 242402 (2014).